\documentclass[journal]{IEEEtran}

\usepackage{color}
\usepackage{subfigure}
\usepackage{graphics} 
\usepackage{epsfig} 
\usepackage{mathptmx} 

\usepackage{amsmath} 
\usepackage{amssymb}  
\usepackage{psfrag}
\usepackage{bm}
\usepackage{amsbsy}
\newtheorem{thm}{Theorem}

\newtheorem{lemma}[thm]{Lemma}

\title{Coding Improves the Optimal Delay-Throughput Trade-offs in Mobile Ad-Hoc Networks:\\
Two-Dimensional I.I.D. Mobility Models\thanks{An earlier version of this paper appeared in the
Proc. of WiOpt, 2007.}\thanks{Research presented here was supported in part by a Vodafone
Fellowship and NSF grant CNS 05-19691.}}

\author{Lei Ying, Sichao Yang and R. Srikant\\
Coordinated Science Lab\\ and\\ Department of Electrical and Computer Engineering\\
University of Illinois at Urbana-Champaign\\
\{lying,syang8,rsrikant\}@uiuc.edu}

\begin{document}
\maketitle
\begin{abstract}
In this paper, we investigate the delay-throughput trade-offs in mobile ad-hoc networks under
two-dimensional i.i.d. mobility models. We consider two mobility time-scales: (i) Fast mobility
where node mobility is at the same time-scale as data transmissions; (ii) Slow mobility where node
mobility is assumed to occur at a much slower time-scale than data transmissions. Given a delay
constraint $D,$ the main results are as follows: (1) For the two-dimensional i.i.d. mobility model
with fast mobiles, the maximum throughput per source-destination (S-D) pair is shown to be
$O\left(\sqrt{D/n}\right),$ where $n$ is the number of mobiles. (2) For the two-dimensional i.i.d.
mobility model with slow mobiles, the maximum throughput per S-D pair is shown to be
$O\left(\sqrt[3]{D/n}\right).$ (3) For each case, we propose a joint coding-scheduling algorithm to
achieve the optimal delay-throughput trade-offs.
\end{abstract}

\section{Notations}
The following notations are used throughout this paper, given non-negative functions $f(n)$ and
$g(n)$:
\begin{enumerate} \item[(1)]
$f(n)=O(g(n))$ means there exist positive constants $c$ and $m$ such that $f(n) \leq cg(n)$ for all
$ n\geq m$.

\item[(2)] $f(n)=\Omega(g(n))$ means there exist positive constants $c$ and $m$ such that $f(n)\geq
cg(n)$ for all $n\geq m.$ Namely, $g(n)=O(f(n))$.

\item[(3)] $f(n)=\Theta(g(n))$ means  that both $f(n)=\Omega(g(n))$ and $f(n)=O(g(n))$ hold.

\item[(4)]$f(n)=o(g(n))$ means that $\lim_{n\rightarrow \infty} f(n)/g(n)=0.$

\item[(5)] $f(n)=\omega(g(n))$ means that $\lim_{n\rightarrow \infty} g(n)/f(n)=0.$ Namely,
$g(n)=o(f(n)).$

\end{enumerate}

\section{Introduction}
The throughput of a random wireless network with $n$ static nodes and $n$ random S-D pairs was
studied by Gupta and Kumar \cite{gupkum00}. They showed that the maximum throughput per S-D pair is
$O(1/\sqrt{n}),$ and proposed a scheduling scheme achieving a throughput of $\Theta(1/\sqrt{n\log
n})$ per S-D pair. The throughput decreases with $n$ because each successful transmission from
source to destination needs to take $\sqrt{n/\log n}$ hops. Later Grossglauser and Tse
\cite{GroTse_01} considered mobile ad-hoc networks, and showed that $\Theta(1)$ throughput per S-D
pair is achievable. The idea is to deliver a packet to its destination only when it is within
distance $\Theta(1/\sqrt{n})$ from the destination. However, packets have to tolerate large delays
to achieve this throughput.

We first review the results for i.i.d. mobility models. Neely and Modiano \cite{NeeMod_05} studied
the i.i.d. mobility model where the positions of nodes are totally reshuffled from one time slot to
another, and showed that the mean delay of Grossglauser and Tse's algorithm is $\Theta(n).$ In the
same paper, they also proposed an algorithm which generates multiple copies of each data packet to
reduce the mean delay. Since more transmissions are required when we generate multiple copies, the
throughput per S-D decreases with the number of copies per data packet. The delay-throughput
trade-off is shown to be $\lambda=\Omega(D/n)$ in \cite{NeeMod_05}, where $\lambda$ is the
throughput per S-D pair, and $D$ is the number of time slots taken to deliver packets from source
to destination.

In \cite{NeeMod_05}, fast mobility is assumed. A different time-scale of mobility, slow mobility,
was considered by Toumpis and Goldsmith in \cite{TouGol_04}, and Lin and Shroff in
\cite{LinShr_04}. For slow mobiles, node mobility is assumed to be much slower than data
transmissions. So the packet size can be scaled down as $n$ increases, and multi-hop transmissions
are feasible in single time slot. The delay-throughput trade-off was shown to be
$\lambda=\Omega\left( \sqrt{D/n} \log n\right)$ in \cite{TouGol_04}. A better trade-off was
obtained in \cite{LinShr_04}, where the maximum throughput per S-D pair for mean delay $D$ was
shown to be $\lambda=O\left(\sqrt[3]{D/n}\log n\right),$ and a scheme was proposed to achieve a
trade-off of $\lambda=\Theta\left(\sqrt[3]{D/\left(n\log^{9/2} n\right)}\right).$

Besides the i.i.d. mobility model, other mobility models have also been studied in the literature.
The random walk model was introduced by El~Gamal \emph{et al} in \cite{GamMamPraSha_04}, and later
studied in \cite{GamMamParSha_06, GamMamParSha_06_1} and \cite{ShaMazShr_06}. In
\cite{GamMamParSha_06} and \cite{GamMamParSha_06_1}, the throughput per S-D pair is shown to be
$\Theta(1/\sqrt{n\log n})$ for $D=O(\sqrt{n/\log n}),$  and $\Theta(D/n)$ for
$D=\Omega(\sqrt{n/\log n}),$ where \cite{GamMamParSha_06} focused on the slow mobility and
\cite{GamMamParSha_06_1} focused on the fast mobility. Other mobility models, like Brownian motion,
one dimensional mobility, and hybrid random walk models have been studied in
\cite{LinShaMazShr_06}, \cite{DigGroTse_02}, \cite{GamMamParSha_06_2} and \cite{ShaMazShr_06}.

Although the delay-throughput trade-off has been widely studied for various mobility models, the
optimal delay-throughput trade-off has not yet been established except for two cases of mobility
models \cite{GamMamParSha_06}, \cite{GamMamParSha_06_1}, \cite{LinShaMazShr_06}. In this paper, we
investigate ad-hoc networks with the two-dimensional i.i.d. mobility. Our main results are as
follows:
\begin{enumerate}
\item[(1)] For the two-dimensional i.i.d. mobility model with fast mobiles, we show that the
maximum throughput per S-D pair is $O\left(\sqrt{D/n}\right)$ under a delay constraint $D.$ A joint
coding-scheduling algorithm is presented to achieve the maximum throughput for $D$ is both
$\omega\left(\sqrt[3]{n}\right)$ and $o(n).$

\item[(2)] For the two-dimensional i.i.d. mobility model with slow mobiles, we first prove that the
maximum throughput per S-D pair is $O\left(\sqrt[3]{D/n}\right)$ given a delay constraint $D.$ Then
we propose another joint coding-scheduling algorithm to achieve the maximum throughput for $D$ is
both $\omega(1)$ and $o(n).$ In both case (1) and (2), we need a lower bound on delay to ensure
decodability of packets with high probability for large $n.$
\end{enumerate}

The above results can be extended to other mobility models as shown in a companion paper
\cite{YinYanSri_06}.

We also would like to mention that there is a very recent result by Ozgur, Leveque, and Tse
\cite{OzgLevTse_06} where they showed a throughput of $\Theta(1)$ per S-D pair is achievable using
node cooperation and MIMO communication; see also the earlier paper by Aeron and Saligrama in
\cite{AerSal_06}.  These schemes require sophisticated signal processing techniques, not considered
in this paper.

The remainder of the paper is organized as follows: In Section \ref{sec: model}, we introduce the
communication and mobility model. Main results along with some intuition into them are presented in
Section \ref{sec: K}. Then we analyze the two-dimensional i.i.d. mobility models with fast mobiles
in Section \ref{sec: IID_STSM} and slow mobiles in Section \ref{sec: IID_TSSM}. Finally, the
conclusions is given in Section \ref{sec: conl}. In the appendix, we collect some results that are
frequently used in the paper.

\section{Model}
\label{sec: model} In this section, we first present the mobility and wireless interference models
used in this paper. Then the definitions of delay and throughput are provided.

\noindent{\bf Mobile Ad-Hoc Network Model:} Consider an ad-hoc network where wireless mobile nodes
are positioned in a unit square. Assuming the time is slotted, we study the two-dimensional i.i.d.
mobility model in this paper, which was introduced in \cite{NeeMod_05} and defined as follows:
\begin{enumerate}
\item[(i)] There are $n$ wireless mobile nodes positioned on a unit square. At each time slot, the
nodes are uniformly, randomly positioned in the unit square.

\item[(ii)] The node positions are independent of each other, and independent from time slot to
time slot. So the nodes are totally reshuffled at each time slot.

\item[(iii)] There are $n$ S-D pairs in the network. Each node is both a source and a destination.
Without loss of generality, we assume that the destination of node $i$ is node $i+1,$ and the
destination of node $n$ is node $1.$
\end{enumerate}

\noindent{\bf Communication Model:} We assume the protocol model introduced in \cite{gupkum00} in
this paper. Let $\hbox{dist}(i,j)$ denote the Euclidean distance between node $i$ and node $j,$ and
$r_i$ to denote the transmission radius of node $i.$ A transmission from node $i$ can be
successfully received at node $j$ if and only if following two conditions hold:
\begin{enumerate}
\item[(i)] $\hbox{dist}(i,j)\leq r_i;$

\item[(ii)] $\hbox{dist}(k,j)\geq (1+\Delta) \hbox{dist}(i,j)$ for each node $k\not=i$ which
transmits at the same time, where $\Delta$ is a protocol-specified guard-zone to prevent
interference.
\end{enumerate}
We further assume that at each time slot, at most $W$ bits can be transmitted in a successful
transmission.

\noindent{\bf Time-Scale of Mobility:}  Two time-scales of mobility are considered in this paper.
\begin{enumerate}
\item[(1)] Fast mobility: The mobility of nodes is at the same time-scale as the data transmission,
so $W$ is a constant independent of $n$ and only one-hop transmissions are feasible in single time
slot.

\item[(2)] Slow mobility: The mobility of nodes is much slower than the wireless transmission, so
$W\gg n.$ Under this assumption, the packet size can be scaled as $W/H(n)$ for $H(n)=O(n)$ to
guarantee $H(n)$-hop transmissions are feasible in single time slot.

\end{enumerate}

\noindent{\bf Delay and Throughput:} We consider hard delay constraints in this paper. Given a
delay constraint $D,$ a packet is said to be successfully delivered if the destination obtains the
packet within $D$ time slots after it is sent out from the source.

Let $\Lambda_i[T]$ denote the number of bits successfully delivered to the destination of node $i$
in time interval $[0, T].$ A throughput of $\lambda$ per S-D pair is said to be feasible under the
delay constraint $D$ and loss probability constraint $\epsilon>0$ if there exists $n_0$ such that
for any $n\geq n_0,$ there exists a coding/routing/scheduling algorithm with the property that each
bit transmitted by a source is received at its destination with probability at least $1-\epsilon,$
and
\begin{equation} \lim_{T\rightarrow \infty}\Pr\left( \displaystyle
\frac{\Lambda_i[T]}{T}\geq \lambda, \hbox{ }\forall \hbox{ } i\right)=1. \label{eq: throughput}
\end{equation}

\section{Main Results and Some Intuition}
\label{sec: K} Recall that our objective is to maximize throughput in a wireless network subject to
a delay constraint and a wireless interference constraint.  More precisely, the constraints can be
viewed as follows:
\begin{enumerate}
\item[(1)] Wireless interference: Throughput is limited due to the fact that transmissions
interfere with each other.

\item[(2)] Mobility: A packet may not be delivered to its destination before the delay deadline
since neither the packet's source nor any relay node may get close enough to the destination.
\end{enumerate}
In this section, we present some heuristic arguments to obtain an upper bound on the maximum
throughput subject to these two constraints and derive the key results of the paper. While the
heuristics are far from precise derivations of the optimal delay-throughput trade-offs, they may be
useful to the reader in understanding the main results. In addition, the heuristic arguments
provide the right order for the ``hitting distance" (to be defined later) which plays a critical
role in the optimal scheme used to achieve the delay-throughput trade-offs.

\begin{figure}[hbt] \psfrag{P}{$\scriptstyle P_e=(1-\pi L^2)^D$}
\psfrag{R}{$\scriptstyle R=\frac{1}{c_1 L^2 n}$}
\centering{\epsfig{file=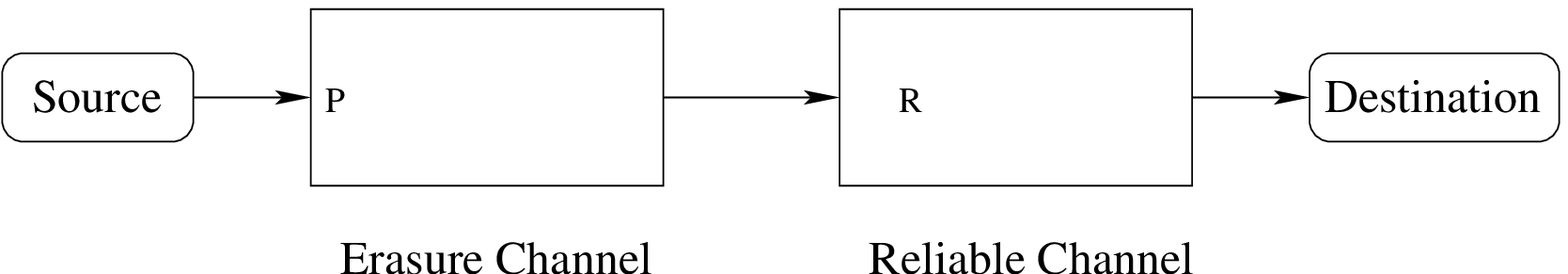,width=3.5in}} \caption{Virtual-channel Representation for the
Tow-Dimensional I.I.D. Mobility Model with Fast Mobiles} \label{fig: virtual2}
\end{figure}

Consider the two-dimensional i.i.d. mobility model with fast mobiles. We say that a packet
\emph{hits} its destination at time slot $t$ if the distance between the packet and its destination
is less than or equal to $L.$ Under the two-dimensional i.i.d. mobility model, a packet hits its
destination with probability $\pi L^2$ at each time slot. So given a delay constraint $D,$ the
probability that a packet hits its destination in one of $D$ time slots is
$$1-\left(1-\pi L^2\right)^D.$$ Furthermore under the fast mobility, only one-hop transmissions are feasible at each time slot.
So the transmission radius needs to be at least $L$ to deliver packets to the destinations when
their distance is $L.$ Assume all nodes use a common transmission radius $L$ and that all nodes
wish to transmit at each time slot, then each node has $1/(c_1 nL^2)$ fraction of time to transmit,
and the throughput per S-D pair is no more than $1/(c_1 nL^2)$ where $c_1$ is a positive constant
independent of $n.$ Thus the network can be regarded as a system where there are two virtual
channels between each S-D pair as in Figure \ref{fig: virtual2}. The packets are first sent over
the erasure channel with erasure probability
$$P_e=(1-\pi L^2)^D,$$ and then over the reliable channel with rate $$R=\frac{1}{c_1 L^2 n}$$ bits per
time slot. Each source can transmit at most $W$ bits per time slot on average. So in this virtual
system, the maximum throughput of a S-D pair is
\begin{eqnarray*}\lambda&=&\max_{L} \min
\left\{W\left(1-\left(1-\pi L^2\right)^D\right), \frac{1}{c_1 L^2 n}\right\} \\
&=&\sqrt{\frac{ \pi W D}{c_1 n}} ,
\end{eqnarray*} and the corresponding optimal hitting distance $L^*=b_1/\sqrt[4]{nD}$ where $b_1=\sqrt[4]{c_1\pi W}.$

To achieve this throughput, we first need to use the optimal $L.$ Furthermore, \emph{a coding
scheme achieving the capacity of the erasure channel} is needed. Since the erasure probability is
determined by $L$ and $D,$ which are different under different delay constraints, rate-less codes
become a reasonable choice. The key idea in this paper is to encode data packets using Raptor
codes, which are near optimal rate-less codes with low complexity. We also note that the idea of
using coding to improve reliability of packet delivery has also been considered by Shah and
Shakkottai in \cite{ShaSha_06} for ad hoc sensor networks in a different context. Our first result
is as follows.

\noindent{\bf Main Result 1:} Under the two-dimensional i.i.d. mobility model with fast mobiles,
the throughput per S-D pair is $\lambda=O\left(\sqrt{D/n}\right)$ given a delay constraint $D.$ For
$D$ is both $\omega(\sqrt[3]{n})$ and $o(n),$ this throughput can be achieved using a joint
coding-scheduling algorithm.

Note that the heuristic arguments leading up to the above result have many flaws. For example, it
suggests that one can wait for the source to hit the destination to deliver the packet. In reality,
such a scheme will not work since we deliver only one packet to the destination during each
encounter between the S-D pair. Thus other packets at the source which are not delivered may
violate their delay constraints. This problem in the heuristic argument is due to the fact that it
assumes that we have an independent erasure channel for each packet despite the fact that the
transmitting node is the same source. Despite the flaws, the heuristic argument surprisingly
captures the delay-throughput trade-off and the optimal hitting distance correctly up to the right
order. In practice, the bound is achievable by exploiting the broadcast nature of the wireless
channel to transmit each packet to several relay nodes and allowing relay nodes to independently
attempt to deliver the packet the destination.

\begin{figure}[hbt]
\psfrag{P}{$\scriptstyle P_e=(1-\pi L^2)^D$} \psfrag{R}{$\scriptstyle R=\frac{1}{c_2 L\sqrt{n}}$}
\centering{\epsfig{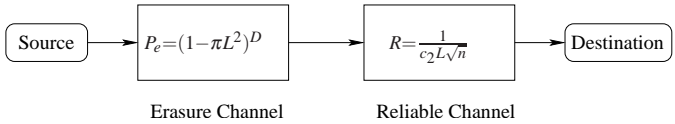}} \caption{Virtual-channel Representation for the
Two-Dimensional I.I.D. Mobility Model with Slow Mobiles} \label{fig: virtual}
\end{figure}

Next consider the two-dimensional i.i.d. mobility model with slow mobiles. Since multi-hop
transmissions are feasible at each time slot, using  a precise version of the result
\cite{gupkum00} which was obtained in \cite{FraDouTseThi_05}, the maximum throughput per S-D pair
under the slow mobility assumption is
$$\frac{1}{c_2 L\sqrt{n}},$$ where $c_2$ is a positive constant independent of $n.$ We provide a crude version of the
argument from \cite{gupkum00} here for ease of readability. Suppose each node uses a transmission
radius $r$ and the distance between a S-D pair is $L,$ then each bit has to travel $L/r$ hops. The
number of bit-hops needed to satisfy a throughput requirement of $\lambda$ bits/slot/node in $T$
slots is $\lambda L T/r.$ Due to the interference model, the number of simultaneous transmissions
possible in one time slot is $1/(\tilde{c}_2r^2)$ for some constant $\tilde{c}_2.$ Thus we need
$$\frac{n\lambda L T}{r}\leq \frac{T}{\tilde{c}_2 r^2},$$ or $$\lambda\leq \frac{1}{\tilde{c}_2 L r n}.$$
Intuitively, since the total area is $1$ and the number of nodes is $n,$ the smallest radius of
transmission that can be used while ensuring connectivity is given by $n\pi r^2=1,$ so
$$\lambda\leq \frac{1}{\tilde{c}_2 L\sqrt{\pi n}}.$$ That this is indeed achievable in an order sense is proved in
\cite{FraDouTseThi_05}, and therefore, we take $\lambda$ to be $1/(c_2 L \sqrt{n})$ where
$c_2=\sqrt{\pi}\tilde{c}_2.$ Then the virtual channels between a S-D pair are as depicted in Figure
\ref{fig: virtual}. In this virtual system, the maximum throughput of a S-D pair is
\begin{eqnarray*}\lambda&=&\max_{L} \min
\left\{W\left(1-\left(1-\pi L^2\right)^D\right), \frac{1}{c_2 L\sqrt{n}}\right\} \\
&=&\sqrt[3]{\frac{ \pi W D}{c_2^2 n}},
\end{eqnarray*} and the optimal hitting distance $L^*=b_2/\sqrt[6]{n D^2}$ where $b_2=\sqrt[3]{c_2\pi W}.$
This throughput can also be achieved using a joint coding-scheduling scheme. The main result is
summarized as follows.

\noindent{\bf Main Result 2:} Under the two-dimensional i.i.d. mobility model with slow mobiles,
the throughput per S-D pair is $\lambda=O\left(\sqrt[3]{D/n}\right)$ given a delay constraint $D.$
This throughput can be achieved using a joint coding-scheduling scheme when $D$ is both $\omega(1)$
and $o(n).$

As stated before, the crude virtual channel representation used in this section surprisingly yields
the correct results. However, they do not form the basis of the proofs in the rest of the paper.
Several assumptions have been made in deriving the virtual channel representation:
\begin{enumerate}
\item[(i)] The hitting events for various packets are assumed to independent which is difficult to
ensure since the same node may act as a relay for multiple packets.

\item[(ii)] It assumes a fixed hitting distance which is not reasonable to obtain an upper bound on
the throughput. An upper bound must be scheme-independent.
\end{enumerate}
In view of these limitations, we use the virtual channel model to only provide some insight into
the results and the hitting distance we should use in the achievable algorithms, but rigorous
proofs of the main results are provided in subsequent sections.

\section{Two-Dimensional I.I.D. Mobility Model, Fast Mobiles}
\label{sec: IID_STSM}

In this section, we investigate the two-dimensional i.i.d. mobility model with fast mobiles.
Assuming that all mobiles have wireless communication and coding capability, we investigate the
maximum throughput the network can achieve by using relaying and coding to recover packet loss as
discussed in the heuristic arguments. Given a delay constraint $D,$ we will first prove that the
maximum throughput per S-D pair which can be supported by the network is
$O\left(\sqrt{D/n}\right).$ Then a joint coding-scheduling scheme will be proposed to achieve the
maximum throughput when $D$ is both $\omega(\sqrt[3]{n})$ and $o(n).$

\subsection{Upper Bound}
In this subsection, we show the maximum throughput the network can support without network coding,
i.e., under the following assumption.

\noindent{\bf Assumption 1:} Packets destined for different nodes cannot be encoded together.
Further, we assume that coding is only used to recover from erasures and not for data compression.
Specifically we assume that at least $k$ coded packets are necessary to recover $k$ data packets,
where all packets (coded or uncoded) are assumed to be of the same size.

Assumption $1$ is the only significant restriction imposed on coding/routing/scheduling schemes. We
also make the following assumption.

\noindent{\bf Assumption 2:} A new coded packet is generated right before the packet is sent out.
The node generating the coded packet does not store the packet in its buffer.

Assumption 2 is not restrictive since the information contained in the new packet is already
available at the node.

\noindent{\bf Assumption 3:} Once a node receives a packet (coded or uncoded), the packet is not
discarded by the node till its deadline expires.

Assumption 3 is not restrictive since we are studying an upper bound on the throughput in this
section.

Next we introduce following notations which will be used in our proof.
\begin{itemize}
\item $b:$ Index of a bit stored in the network. Bit $b$ could be either a bit of a data packet or
a bit of a coded packet. If a node generates a copy of a packet to be stored in another node, then
the bits in the copy are given different indices than the bits in the original packet.

\item $d_b:$ The destination of bit $b.$

\item $c_b:$ The node storing bit $b.$

\item $t_b:$ The time slot at which bit $b$ is generated.

\item $S_{b}:$ If bit $b$ is delivered to its destination, then $S_{b}$ is the transmission radius
used to deliver $b.$

\item ${\cal R}[T]:$ The set of all bits stored at relay nodes at time slot $T.$ We do not include
bits that are still in their source node in defining ${\cal R}[T].$

\item $\Lambda[T]:$ $\Lambda[T]=\sum_{i=1}^n \Lambda_i[T].$
\end{itemize}

Assume that the delay constraint is $D,$ and a data packet is processed by the source node at time
slot $t_p.$ Then the data packet is said to be active from time slot $t_p$ to $t_p+D-1.$ A bit $b$
is said to be active if at least one data packet encoded into the packet containing bit $b$ has not
expired. It is easy to see that any bit expires at most $D$ time slots after the bit is generated.
Also a bit is said to be good if it is active when delivered to its destination. Now let
$\tilde{\Lambda}[T]$ denote the number of good bits delivered to destinations in $[0, T].$ Without
loss of generality, we assume good bits are indexed from $1$ to $\tilde{\Lambda}[T].$ Note that
expired bits might help decode good source bits but would not contribute to the total throughput,
so we have \begin{eqnarray}\tilde{\Lambda}[T]\geq \Lambda [T],\label{eq: la}\end{eqnarray} where
$\Lambda[T]$ is the number of good source bits successfully recovered at destinations.

Next we present three fundamental constraints. In the following lemma, inequalities (\ref{eq:
c1_d}) and (\ref{eq: c11_d}) hold since the total number of bits transmitted or received in $T$
time slots cannot exceed $nWT.$ Inequality (\ref{eq: c2_d}) holds since under the protocol model,
discs of radius $\Delta r_i/2$ around the receivers should be mutually disjoint from each other.
\begin{lemma}
For any mobility model, the following inequalities hold,
\begin{eqnarray}
\tilde{\Lambda}[T] &\leq& nWT \label{eq: c1_d}\\
|{\cal R}[T]| &\leq& nWT \label{eq: c11_d}\\
\sum_{b=1}^{\tilde{\Lambda}[T]}\frac{\Delta^2}{16} \left(S_{b}\right)^2 &\leq & \frac{WT}{\pi},
\label{eq: c2_d}
\end{eqnarray}
where $|{\cal R}[T]|$ is the cardinality of the set ${\cal R}[T].$ \label{lem: c_d}
\end{lemma}
\begin{proof}
Since each node can transmit at most $W$ bits per time slot, the total number of bits transmitted
in $T$ time slots is less than $nWT$ which implies inequalities (\ref{eq: c1_d}) and (\ref{eq:
c11_d}). Inequality (\ref{eq: c2_d}) was proved in \cite{AgaKum_04}.
\end{proof}

We first consider the scenario where packet relaying is not allowed, i.e., packets need to be
directly transmitted from sources to destinations. In the following lemma, we show that the
throughput in this case is at most $\Theta(1/\sqrt{n})$ even without the delay constraint.
\begin{lemma}
Consider the two-dimensional i.i.d. mobility model with fast mobiles. Suppose that packets have to
be directly transmitted from sources to destinations, then \begin{eqnarray}
\frac{8\sqrt{2}}{\Delta}WT\sqrt{n} \geq E\left[{{\Lambda}[T]}\right]. \label{eq: norelay}
\end{eqnarray} \label{lem: norelay}
\end{lemma}
\begin{proof}
First from the Cauchy-Schwarz inequality and inequality (\ref{eq: c2_d}), we have
\begin{eqnarray*}
\left(\sum_{b=1}^{\tilde{\Lambda}[T]} S_{b}\right)^2 &\leq&\left(\sum_{b=1}^{\tilde{\Lambda}[T]}
1\right)\left(\sum_{b=1}^{ \tilde{\Lambda}[T]}\left(S_{b}\right)^2\right)\\
&\leq & \tilde{\Lambda}[T]\frac{16 W T}{\pi\Delta^2} ,
\end{eqnarray*} which implies
\begin{eqnarray}
E\left[\sum_{b=1}^{\tilde{\Lambda}[T]} S_{b} \right]\leq \left(\sqrt{\frac{16 W
T}{\pi\Delta^2}}\right)E\left[\sqrt{\tilde{\Lambda}[T]}\right]. \label{eq: up}
\end{eqnarray}
This gives an upper-bound on the expected distance travelled. Next we bound the total number of
times that each mobile gets within a distance $L$ of its destination for $L\in[0, 1/2].$ From the
i.i.d. mobility assumption, we have that for any $i,$ $j$ and $t,$
$$\Pr\left(\hbox{dist}(i,j)(t)\leq L\right)=\pi L^2,$$ which implies
$$E\left[\sum_{t=1}^T\left(\sum_{i=1}^{n}  1_{\hbox{dist}(i,((i+1)\bmod{n}))(t)\leq L}\right)\right]= \pi L^2 n T.$$
Since at most $W$ bits can be transmitted at each time slot, we further have
\begin{eqnarray*}
\sum_{b=1}^{\tilde{\Lambda}[T]} 1_{S_b\leq L}\leq W \sum_{t=1}^{T}\sum_{i=1}^{n}
1_{\hbox{dist}(i,((i+1)\bmod{n}))(t)\leq L}.
\end{eqnarray*} Taking expectation on both sides of above inequality, we obtain
\begin{eqnarray}
E\left[\tilde{\Lambda}[T]\right]-E\left[\sum_{b=1}^{\tilde{\Lambda}[T]} 1_{S_b> L}\right]\leq W \pi
L^2 n T. \label{eq: lo}
\end{eqnarray}

Now using Jensen's inequality and inequalities (\ref{eq: up}) and (\ref{eq: lo}), we can conclude
that
\begin{eqnarray}
\sqrt{\frac{16 W T}{\pi\Delta^2}E\left[\tilde{\Lambda}[T]\right]}&\geq&\left(\sqrt{\frac{16 W
T}{\pi\Delta^2}}\right)E\left[\sqrt{\tilde{\Lambda}[T]}\right]\nonumber\\
&\geq& E\left[\sum_{b=1}^{\tilde{\Lambda}[T]} S_{b} \right]\nonumber\\
&\geq& L E\left[\sum_{b=1}^{\tilde{\Lambda}[T]}
1_{S_{b}> L} \right]\nonumber\\
&\geq& L\left(E\left[\tilde{\Lambda}[T]\right]-W\pi L^2 nT\right).\label{eq: bd}
\end{eqnarray} Note that inequality (\ref{eq: bd}) holds for any $L\in [0, 1/2].$ We choose
\begin{eqnarray*}
L^*=\sqrt{\frac{E[\tilde{\Lambda}[T]]}{2 \pi nWT}},
\end{eqnarray*} which is less than $1/2$ since $\tilde{\Lambda}[T]\leq nWT.$  Substituting $L^*$ into inequality (\ref{eq: bd}), we
have
\begin{eqnarray*}
\sqrt{\frac{16 W T}{\pi\Delta^2}E\left[\tilde{\Lambda}[T]\right]} &\geq& \frac{1}{2} L^*
E\left[{\tilde{\Lambda}[T]}\right],
\end{eqnarray*} which implies that
\begin{eqnarray*}
\frac{8\sqrt{2}}{\Delta}WT\sqrt{n} &\geq& E\left[{\tilde{\Lambda}[T]}\right].
\end{eqnarray*}
The lemma then follows from inequality (\ref{eq: la}).
\end{proof}

Next we investigate the maximum throughput the network can support using coding/routing/scheduling
schemes. We have obtained an upper bound on the number of bits directly transmitted from sources to
destinations in Lemma \ref{lem: norelay}. To bound the maximum throughput with relaying, we will
calculate the number of bits transmitted from relays to destinations in the following analysis.

\begin{thm} Consider the two-dimensional i.i.d. mobility model with fast mobiles, and assume that Assumption 1-3 hold. Then given a delay constraint $D,$
we have that
\begin{eqnarray}
\frac{8\sqrt{2} WT}{\Delta}\sqrt{n}\left(\sqrt{D}+1\right)\geq E\left[{\Lambda}[T]\right].
\label{eq: nrt}
\end{eqnarray}
\label{thm: upper_d}
\end{thm}
\begin{proof} In the proof of the theorem, we treat active bits at relays and active bits at sources differently since we can bound
the number of active bits at relays using inequality (\ref{eq: c11_d}), while the number of active
bits at sources could be larger. Let $\tilde{\Lambda}^r[T]$ denote the number of good bits
delivered directly from relays to destinations in$[0,T].$ Without loss of generality, we assume
these good bits are indexed from $1$ to $\tilde{\Lambda}^r[T].$ Similar to inequality (\ref{eq:
up}), we first have
\begin{eqnarray}
E\left[\sum_{b=1}^{\tilde{\Lambda}^r[T]} S_{b} \right]\leq \left(\sqrt{\frac{16 W
T}{\pi\Delta^2}}\right)E\left[\sqrt{\tilde{\Lambda}^r[T]}\right]. \label{eq: up_ks}
\end{eqnarray}
Let $\tilde{L}_{b}$ denote the minimum distance between node $d_b$ and node $c_b$ from time slot
$t_b$ to time slot $t_b+D-1,$ i.e.,
$$\tilde{L}_b=\min_{t_b\leq t\leq t_b+D-1} \hbox{dist}(d_b, c_b)(t).$$ Then for any $L\in[0, 1/2]$ and any bit $b\in{\cal
R}[T],$ we have
$$\Pr\left(\tilde{L}_b\leq L\right)= 1-(1-\pi L^2)^D\leq \pi L^2
D,$$ which implies
$$E\left[\sum_{b\in {\cal R}[T]} 1_{\tilde{L}_b\leq L} \right]\leq  nWT  \pi L^2 D.$$  Furthermore, we have
\begin{eqnarray*}
\sum_{b=1}^{\tilde{\Lambda}^r[T]} 1_{S_b\leq L}\leq \sum_{b\in {\cal R}[T]} 1_{\tilde{L}_b\leq L },
\end{eqnarray*} which implies that
\begin{eqnarray}
E\left[\sum_{b=1}^{\tilde{\Lambda}^r[T]} 1_{S_b\leq L}\right]\leq nWT  \pi L^2 D. \label{eq: up_3}
\end{eqnarray}
Thus we can conclude that
\begin{eqnarray}
E\left[\sum_{b=1}^{\tilde{\Lambda}^r[T]} S_b \right]&\geq&E\left[\sum_{b=1}^{\tilde{\Lambda}^r[T]}
S_b 1_{S_b>
L}\right]\nonumber\\
&\geq& L E\left[\sum_{b=1}^{\tilde{\Lambda}^r[T]} 1_{S_b> L}\right]\nonumber\\
&\geq& L\left(E[\tilde{\Lambda}^r[T]]-E\left[\sum_{b=1}^{\tilde{\Lambda}^r[T]} 1_{S_b\leq L}\right]\right)\nonumber\\
&\geq& L E[\tilde{\Lambda}^r[T]]-n WT\pi L^3 D,\label{eq: up_2}
\end{eqnarray} where the last inequality follows from inequality (\ref{eq: up_3}).

Now using Jensen's inequality and inequalities (\ref{eq: up_ks}) and (\ref{eq: up_2}), we have that
for any $L\in [0, 1/2],$
\begin{eqnarray}
\sqrt{\frac{16 W T}{\pi\Delta^2}E\left[\tilde{\Lambda}^r[T]\right]}\geq L E[\tilde{\Lambda}^r[T]]-n
WT\pi L^3 D. \label{eq: upp}
\end{eqnarray} Substituting $$L^*=\sqrt{\frac{E\left[\tilde{\Lambda}^r[T]\right]}{2 n
WT\pi D}}$$ into inequality (\ref{eq: upp}), we can conclude that \begin{eqnarray}\frac{8\sqrt{2}
WT}{\Delta}\sqrt{nD}\geq E\left[\tilde{\Lambda}^r[T]\right].\label{eq: upper_d}\end{eqnarray} The
theorem follows from inequalities (\ref{eq: norelay}), (\ref{eq: upper_d}) and (\ref{eq: la}).
\end{proof}

From Theorem \ref{thm: upper_d}, we can conclude that the throughput per S-D is $O(\sqrt{D/n})$
given a delay constraint $D.$

\subsection{Joint Coding-Scheduling Algorithm}
In Section \ref{sec: K}, we motivated the need to first encode data packets. In this subsection, we
use Raptor codes and propose  a joint coding-scheduling scheme to achieve the maximum throughput
obtained in Theorem \ref{thm: upper_d}.

Motivated by the heuristic argument in Section \ref{sec: K}, we divide the unit square into square
cells with each side of length equal to $1/\sqrt[4]{nD},$ which is of the same order as the optimal
hitting distance.  In our scheme, we will allow final delivery of a packet to its destination only
when a relay carrying the packet is in the same cell as the destination. Thus, a packet is
delivered only when the relay and destination are within a distance of $\sqrt{2}/\sqrt[4]{nD},$
which is also the same as the hitting distance calculated in Section \ref{sec: K} except for a
constant factor which does not play a role in the order calculations. The mean number of nodes in
each cell will be denoted by $M$ and is equal to $\sqrt{n/D}.$ The transmission radius of each node
is chosen to be $\sqrt{2}/\sqrt[4]{nD}$ so that any two nodes within a cell can communicate with
each other. This means that, given the interference constraint, two nodes in a cell can communicate
if all nodes in cells within a fixed distance from the given cell stay silent. Each time slot is
further divided into $C$ mini-slots and each cell is guaranteed to be active in at least one
mini-slot within each time slot. Assume $C=9.$ The reason we use nine mini-slots is that if a node
in a cell is active, then no other nodes in any of its neighboring eight cells can be active, but
nodes outside this neighborhood can be active. Further, we denote the packet size to be $W/(2C)$ so
that two packets can be transmitted in each mini-slot.

A cell is said to be a \emph{good cell} at time $t$ if the number of nodes in the cell is between
$9M/10+1$ and $11M/10.$ We also define and categorize packets into four different types.
\begin{itemize}
\item Data packets: There are the uncoded data packets that have to be transmitted by the sources
and received by the destinations.

\item Coded packets: Packets generated by Raptor codes. We let $(i, k)$ denote the $k^{\rm th}$
coded packet of node $i.$

\item Duplicate packets: Each coded packet could be broadcast to other nodes to generate multiple
copies, called duplicate packets. We let $(i,k,j)$ denote a copy of $(i,k)$ carried by node $j,$
and $(i,k, J)$ to denote the set of all copies of coded packet $(i,k).$

\item Deliverable packets: Duplicate packets that happen to be within distance $L$ from their
destinations.
\end{itemize}

We now describe our coding/scheduling algorithm.

\noindent{\bf Joint Coding-Scheduling Scheme I:} We group every $6D$ time slots into a super time
slot. At each super time slot, the nodes transmit packets as follows.
\begin{enumerate}

\item[(1)] {\bf Raptor Encoding:} Each source takes $6D/(25M)$ data packets, and uses Raptor codes
to generate $D/M$ coded packets.

\item[(2)]  {\bf Broadcasting:} This step consists of $D$ time slots. At each time slot, the nodes
executes the following tasks:
\begin{enumerate}
\item[(i)] In each good cell, one node is randomly selected. If the selected node has not already
transmitted all of its $D/M$ coded packets, then it broadcasts a coded packet that was not
previously  transmitted to $9M/10$ other nodes in the cell during the mini-slot allocated to that
cell. Recall that our choice of packet size allows one node in every good cell to transmit during
every time slot.

\item[(ii)] All nodes check the duplicate packets they have. If more than one duplicate packets
have the same destination, select one at random to keep and drop the others.

\end{enumerate}

\item[(3)]{\bf Receiving:} This step consists of $5D$ time slots.  At each time slot, if a cell
contains no more than two deliverable packets, the deliverable packets are delivered to their
destinations using one-hop transmissions during the mini-slot allocated to that cell. At the end of
this step, all undelivered packets are dropped. The destinations decode the received coded packets
using Raptor decoding.
\end{enumerate}

Note that in describing the algorithm, we did not account for the delays in Raptor encoding and
decoding. However, Raptor codes have linear encoding and decoding complexity. Hence, even if these
delays are taken into account, our order results will not change.

\begin{thm}
Consider Joint Coding-Scheduling Algorithm I. Suppose $D$ is both $\omega(\sqrt[3]{n})$ and $o(n),$
and the delay constraint is $6D.$ Then given any $\epsilon>0,$ there exists $n_0$ such that for any
$n\geq n_0,$ every data packet sent out can be recovered at the destination with probability at
least $1-\epsilon,$ and furthermore
\begin{eqnarray}\lim_{T\rightarrow \infty} \Pr\left(\frac{\Lambda_i[T]}{T}\geq
\frac{9W}{500C}\sqrt{\frac{D}{n}} \hbox{ } \forall i\right)=1.\label{eq: th}
\end{eqnarray}
\label{thm: ESTM}
\end{thm}
\begin{proof}
Let $t_s$ denote the $t_s^{\rm th}$ super time slot. For each super time slot, the proof will show
that the following events happen with high probability.

\noindent{\bf Broadcasting:} At least $16D/(25M)$ coded packets from a source are successfully
duplicated after the broadcasting step with high probability, where a coded packet is said to be
successfully duplicated if the packet is in at least $4M/5$ distinct relay nodes. Letting
$A_i[t_s]$ denote the number of coded packets which are successfully duplicated in super time slot
$t_s,$ we will first show that there exists $n_1$ such that for any $n\geq n_1,$
\begin{eqnarray}\Pr\left(A_i[t_s]\geq \frac{16}{25}\frac{D}{M}\right)\geq
1-3e^{-\frac{D}{500M}}. \label{eq: cdt1_2}
\end{eqnarray}

\noindent{\bf Receiving:} At least $7D/(25M)$ distinct coded packets from a source are delivered to
its destination after the receiving step with high probability. Letting $B_i[t_s]$ denote the
number of distinct coded packets delivered to destination $i+1$ in super time slot $t_s,$ we will
show there exists $n_2$ such that for all $n\geq n_2,$
\begin{eqnarray}
\Pr\left(\left.B_i[t_s]\geq \frac{7}{25} \frac{D}{M}\right| A_i[t_s]\geq
\frac{16}{25}\frac{D}{M}\right)\geq 1-2 e^{-\frac{D}{180M}}.\label{eq: cdt2_2}
\end{eqnarray}

\noindent{\bf Decoding:} The $6D/25M$ data packets from a source are recovered with high
probability. Letting ${\cal E}_i[t_s]$ denote the event such that all $6D/(25M)$ data packets are
fully recovered, we will show that
\begin{eqnarray} \Pr\left({\cal E}_i[t_s] \left| B_i[t_s]\geq
\frac{7}{25}\frac{D}{M}\right.\right)\geq 1-\left(\frac{M}{D}\right)^a \label{eq: cdt3_2}
\end{eqnarray} for some  $a>0.$

Recall that $M=\sqrt{n/D}$ and $D=\omega(\sqrt[3]{n}),$ so $$\lim_{n\rightarrow
\infty}\frac{M}{D}=\lim_{n\rightarrow \infty}\frac{\sqrt{n}}{D\sqrt{D}}=0.$$ Combining inequalities
(\ref{eq: cdt1_2})-(\ref{eq: cdt3_2}), we can conclude that for any $\epsilon \leq 1/19,$ there
exists $n_0\geq \max\{n_1, n_2\}$ such that for $n\geq n_0,$
\begin{eqnarray}
\Pr\left( {\cal E}_i[t_s] \right)\geq 1-\epsilon, \label{eq: recover}
\end{eqnarray} which implies that every data packet sent out can be recovered with probability at
least $1-\epsilon.$ Since $1-\epsilon\geq 18/19,$ from the Chernoff bound (see Lemma~\ref{lem: C_B}
provided in the Appendix C for convenience), we can conclude that for $n\geq n_0,$
\begin{eqnarray*}
\Pr\left( \sum_{t_s=1}^{T_s} 1_{{\cal E}_i[t_s]}\geq \frac{9}{10} T_s \right)\geq
1-e^{-\frac{T_s}{800}},\end{eqnarray*} where we choose $\delta=1/20$ in Lemma \ref{lem:
ball-bin_2}. Note that $\sum_{t_s=1}^{T_s} 1_{{\cal E}_i[t_s]} \geq \frac{9}{10} T_s$ implies at
least
$$\frac{9}{10} T_s\times \frac{6D}{25M} \times \frac{W}{2C}= \frac{27W}{250C} \frac{D T_s}{M} = \frac{27W}{250C} D T_s \sqrt{\frac{D}{n}} $$ bits are successfully transmitted from node $i$ to
node $i+1$ in $T_s$ super time slots. Since each super time slot consists of $6D$ time slots, we
can conclude that for $n\geq n_0,$
\begin{eqnarray*}
\Pr\left(\Lambda_i[6DT_s]\geq \frac{27W}{250C} D T_s \sqrt{\frac{D}{n}} \hbox{ }\forall
i\right)\geq 1-ne^{-\frac{T_s}{800}},
\end{eqnarray*} which implies that, for a fixed $n\geq n_0,$
\begin{eqnarray*} \lim_{T\rightarrow\infty} \Pr\left(\frac{\Lambda_i[T]}{T} \geq
\frac{9W}{500C}\sqrt{\frac{D}{n}} \hbox{ }\forall i\right)=1.
\end{eqnarray*}

To complete the proof,  we now show inequalities (\ref{eq: cdt1_2}), (\ref{eq: cdt2_2}) and
(\ref{eq: cdt3_2}).

{\bf Analysis of broadcasting:} Let ${\cal B}_i[t]$ denote the event that node $i$ broadcasts a
coded packet at time slot $t.$ So ${\cal B}_i[t]$ occurs when following two conditions hold:
\begin{enumerate}
\item[(i)] The cell node $i$ in is a good cell;

\item[(ii)] Node $i$ is selected to broadcast.
\end{enumerate}
Since the nodes are uniformly randomly positioned, from the Chernoff bound we have
$$\Pr\left({{\cal B}_i[t]}\right)\geq \frac{10}{11M}\left(1-2e^{-\frac{M}{300}}\right),$$ which
implies that there exists $\tilde{n}_1$ such that for any $n\geq \tilde{n}_1,$ $$\Pr\left({{\cal
B}_i[t]}\right)\geq \frac{8}{9}\frac{1}{M}.$$ Then from the Chernoff bound again, we have
\begin{eqnarray}
\Pr\left(\sum_{t=1}^{D} 1_{{\cal B}_{i}[t]}\geq \frac{4}{5}\frac{D}{M}\right) \geq
1-e^{-\frac{D}{300M}}\label{eq: SB_2}
\end{eqnarray} for $n\geq \tilde{n}_1.$
Thus, with a high probability, more than $4D/(5M)$ coded packets are broadcast, and each broadcast
generates $9M/10$ copies.

Duplicate packets might be dropped at step (ii) of the broadcasting step. We next calculate the
number of duplicate packets of node $i$ left after the broadcasting step.  Assume node $i$
broadcasts $\tilde{D}_i$ coded packets, so $\tilde{D}_i\leq D/M.$  Then the number of duplicate
packets left after the broadcasting step is the same as the number of nonempty bins of following
balls-and-bins problem, where the bins represent the mobile nodes other than node $i,$ and the
balls represent the duplicate packets broadcast from node $i.$

\emph{Balls-and-Bins Problem:} Assume we have $(n-1)$ bins. At each time slot, we select $9 M/10$
bins and drop one ball in each of them. Repeat this $\tilde{D}_i$ times.

Using $N_1$ to denote this number, from Lemma \ref{lem: ball-bin_2} in Appendix C, we have
$$\Pr\left(N_1\geq (1-\delta)(n-1)\tilde{p}_1\right) \geq 1-2e^{-\delta^2 (n-1)\tilde{p}_1/3},$$
where $$\tilde{p}_1=\left(1-e^{-\frac{9 \tilde{D}_i M}{10n-10}}\right).$$ Using the fact
$1-e^{-x}\geq x-x^2/2$ for any $x\geq 0,$ we get
\begin{eqnarray*}
(n-1)\tilde{p}_1&=&(n-1)\left(1-e^{-\frac{9 \tilde{D}_i M}{10n-10}}\right)\\
&\geq& \frac{9\tilde{D}_i M}{10}-\frac{81\tilde{D}_i^2M^2}{100n-100}\\
&\geq& \frac{44}{49}\tilde{D}_i M,
\end{eqnarray*} where the last inequality holds for $n\geq \tilde{n}_2$ for some $\tilde{n}_2$ since $\tilde{D}_i M\leq D=o(n).$
Thus choose $\delta=1/50$ and we can conclude for $n\geq \tilde{n}_2,$
\begin{eqnarray}\Pr\left(\left.N_1\geq \frac{22}{25} \tilde{D}_i M  \right| \sum_{t=1}^{D} 1_{{\cal B}_{i}[t]}=\tilde{D}_i\right)\geq 1-2e^{-\frac{\tilde{D}_i
M}{10000}}. \label{eq: bb1_d}\end{eqnarray}  Recall a coded packet is said to be successfully
duplicated if it has at least $4M/5$ copies at the end of the broadcasting step. Inequality
(\ref{eq: bb1_d}) implies for $n\geq \tilde{n}_2,$
\begin{eqnarray*}\Pr\left(\left. A_i \geq \frac{4}{5} \tilde{D}_i  \right| \sum_{t=1}^{D} 1_{{\cal B}_{i}[t]}=\tilde{D}_i\right)\geq 1-2e^{-\frac{\tilde{D}_i
M}{10000}}, \end{eqnarray*} since otherwise, less than $22\tilde{D}_i M/25$ duplicate packets are
left in the network. Thus we can conclude that for $n\geq \tilde{n}_2,$
\begin{eqnarray}\Pr\left(\left.A_i\geq \frac{16}{25} \frac{D}{M} \right|\sum_{t=1}^{D}
1_{{\cal B}_{i}[t]}\geq \frac{4}{5}\frac{D}{M} \right)\geq 1-2e^{-\frac{D}{20000}}.\label{eq:
R}\end{eqnarray} Letting $n_1=\max\{\tilde{n}_1, \tilde{n}_2\},$ inequality (\ref{eq: cdt1_2})
follows from inequalities (\ref{eq: SB_2}) and (\ref{eq: R}) for $n\geq n_1.$

{\bf Analysis of receiving:} Assume coded packets $\{(i,1), \ldots, \left(i, 16D/(25M) \right)\}$
are successfully duplicated. We let ${\cal D}_{(i,k)}[t]$ denote the event that coded packet
$(i,k)$ is delivered at time slot $t.$ Then ${{\cal D}_{(i,k)}[t]}$ will definitely occur if both
the following conditions hold:
\begin{enumerate}
\item[(i)] One and only one duplicate packet of $(i,k)$ becomes a deliverable packet. Let ${\cal
D}^1_{(i,k)}[t]$ denote this event. Assume the duplicate packet is $(i,k,j),$ i.e., node $j$
contains packet $(i,k).$

\item[(ii)] There are no other deliverable packets in the cell containing node $j$ except packet
$(i,k,j)$ and one possible duplicate packet to node $j$ carried by node $i+1.$ Let ${\cal
D}^2_{(i,k)}[t]$ denote this event.
\end{enumerate}

Note that duplicate packets of node $i$ are carried by different nodes, and their mobilities are
independent. Now assume there are $\tilde{M}_{(i,k)}$ copies of $(i,k)$ in the entire network, then
\begin{eqnarray*}
\Pr\left({\cal D}^1_{(i,k)}[t]\right)& =&\frac{ \tilde{M}_{(i,k)} M}{n}
\left(1-\frac{M}{n}\right)^{\tilde{M}_{(i,k)}-1}.
\end{eqnarray*} Note that $(1-M/n)^{\tilde{M}_{(i,k)}-1}\geq 1-(\tilde{M}_{(i,k)}-1)M/n,$ and $\tilde{M}_{(i,k)}\geq 4M/$ if $(i,k)$ is successfully duplicated.
So for a successfully duplicated packet, there exists $\tilde{n}_3$ such that for any $n\geq
\tilde{n}_3,$
\begin{eqnarray*}
\Pr\left({{\cal D}^1_{(i,k)}[t]}\right)\geq \frac{7 M^2}{10 n}.
\end{eqnarray*}

Suppose we have $\bar{M}$ nodes in the cell containing node $j,$ from the Chernoff bound, we have
$$\Pr\left(\bar{M}\leq \frac{11}{10}M\right)\geq 1-e^{-\frac{M}{300}}.$$ Note that
condition (ii) is equivalent to the following event: Given node $j$ and node $i+1$ in the cell, no
more deliverable packets appear when we put another $\bar{M}-2$ nodes into the cell. Now given $K$
nodes in the cell, the probability that no more deliverable appears when we put another node is at
least
$$\left(1-\frac{2KD}{n-K}\right).$$ This holds due to the following two facts:
\begin{enumerate}
\item[(a)] The new node should not be the destination of any duplicate packets already in the cell
(there are at most $KD$ duplicate packets already in the cell).

\item[(b)] The duplicate packets carried by the new node are not destined for any of the existing
$K$ nodes. Note that each source has no more than $D$ duplicate packets, so there are at most $KD$
nodes which carry the duplicate packet towards the $K$ existing nodes.
\end{enumerate}
Note that $\lim_{n\rightarrow\infty} M=\infty,$ so there exists $\tilde{n}_4$ such that for any
$n\geq \tilde{n}_4,$
\begin{eqnarray*}
Pr\left(\left.{{\cal D}^2_{(i,k)}[t]}\right|{{\cal D}^1_{(i,k)}[t]}\right)
& \geq & \left(1-e^{-\frac{M}{300}}\right) \displaystyle\Pi_{K=2}^{\frac{11M}{10}-1} \left(1-\frac{2KD}{n-K}\right)\\
& \geq & \left(1-e^{-\frac{M}{300}}\right)\left(1-\frac{22MD}{10n-11M}\right)^{\frac{11M}{20}}\\
&\geq &\frac{3}{11}.
\end{eqnarray*}
So we can conclude that for any $n\geq \max\{\tilde{n}_3, \tilde{n}_4\},$
\begin{eqnarray*}
\Pr\left({{\cal D}_{(i,k)}[t]}\right)\geq \frac{21 M^2}{110 n}=\frac{21}{110D},
\end{eqnarray*}
which implies at each time slot, a successfully duplicated packet $(i,k)$ will be delivered with
probability at least $21 /(110 D).$  Note at each time slot, only one coded packet can be delivered
to the destination of node $i.$ So the number of distinct coded packets delivered to the
destination of node $i$ is the same as the number of nonempty bins of following balls-and-bins
problem, where the bins represent the distinct coded packets, the balls represent successful
deliveries, and a ball is dropped in a specific bin means the corresponding coded packet is
delivered to the destination.

\emph{Balls-and-bins Problem:} Suppose we have $16D/(25M)$ bins and one trash can. At each time
slot, we drop a ball. Each bin receives the ball with probability $21/(110 D),$ and the trash can
receives the ball with probability $1-p,$ where
$$p=\frac{21}{110 D}\times \frac{16D}{25M}=\frac{168}{1375}\frac{1}{M}.$$ Repeat this $5D$ times, i.e., $5D$ balls are dropped.

Let $N_2$ denote nonempty bins of the above balls-and-bins problem and choose $\delta=1/6.$ From
Lemma \ref{lem: ball-bin_2} in Appendix C, we have
\begin{eqnarray*}
\Pr\left(N_2\geq \frac{7}{25}\frac{D}{M}\right)\geq 1-2e^{-\frac{D}{180M}},
\end{eqnarray*} and inequality (\ref{eq: cdt2_2}) holds for $n\geq n_2,$ where $n_2=\max\{\tilde{n}_3,
\tilde{n}_4\}.$

{\bf Analysis of decoding:} Inequality (\ref{eq: cdt3_2}) follows from Lemma \ref{lem: RC} on the
error probability of Raptor codes provided in Appendix A.
\end{proof}

\section{Two-Dimensional I.I.D. Mobility Model, Slow Mobiles} \label{sec: IID_TSSM} In this section, we investigate the
two-dimensional i.i.d. mobility model with slow mobiles. Given a delay constraint $D,$ we first
prove the maximum throughput per S-D pair which can be supported by the network is
$O\left(\sqrt[3]{D/n}\right).$ Then a joint coding-scheduling scheme is proposed to achieve the
maximum throughput.

\subsection{Upper Bound} Let $\hat{t}_{b}$ denote the time slot in which bit $b$ is delivered to its destination. Under slow mobility, the
delivery in $\hat{t}_b$ could use multi-hop transmissions, so we further define following
notations:
\begin{itemize}
\item  $H_b:$ The number hops bit $b$ travels in time slot $\hat{t}_b.$

\item $L_b:$ The Euclidean distance bit $b$ travels in time slot $\hat{t}_b.$

\item  $S_b^h:$ The transmission radius used in hop $h$ for $ 1\leq h \leq H_b.$
\end{itemize}
Similar to Lemma \ref{lem: c_d}, we have following results.
\begin{lemma}
For any mobility model, the following inequalities hold,
\begin{eqnarray}
\sum_{b=1}^{\tilde{\Lambda}[T]}\sum_{h=1}^{H_b} 1 &\leq& nWT \label{eq: c1}\\
\sum_{b=1}^{\tilde{\Lambda}[T]}\sum_{h=1}^{H_b}\frac{\Delta^2}{16} \left(S_b^h\right)^2 &\leq &
\frac{WT}{\pi}. \label{eq: c2}
\end{eqnarray}
\label{lem: c} \rightline{$\square$}
\end{lemma}

Similar to the fast mobility cases, we first consider the throughput under the assumption that the
packets can only be delivered to destinations from sources.
\begin{lemma}
Consider the two-dimensional i.i.d. mobility model with slow mobiles. Suppose that packets have to
be directly transmitted to destinations from sources, then
\begin{eqnarray}
\frac{4\sqrt[3]{2}WT}{\sqrt[3/2]{\Delta}}\sqrt[3/2]{n}\geq E[{\Lambda}[T]] \label{eq: nr_s}
\end{eqnarray}
\label{lem: norelay_s}
\end{lemma}
\begin{proof}
First from the Cauchy-Schwartz inequality and Lemma \ref{lem: c}, we have that
\begin{eqnarray*}
\left(\sum_{b=1}^{\tilde{\Lambda}[T]}\sum_{h=1}^{H_b} S_{b}^h\right)^2
&\leq&\left(\sum_{b=1}^{\tilde{\Lambda}[T]}\sum_{h=1}^{H_{b}} 1\right)\left(
\sum_{b=1}^{\tilde{\Lambda}[T]}\sum_{h=1}^{H_{b}}\left(S_{b}^h\right)^2\right)\\
&\leq & W T n \sum_{b=1}^{\tilde{\Lambda}[T]}\sum_{h=1}^{H_{b}} \left(S_{b}^h\right)^2\\
&\leq & \frac{16 W^2 T^2 n}{\pi\Delta^2}, \label{eq: ff}
\end{eqnarray*}
which implies
\begin{eqnarray*}
\frac{4 W T \sqrt{n}}{\Delta \sqrt{\pi} } \geq \sum_{b=1}^{\tilde{\Lambda}[T]} L_{b} \label{eq:
upper}
\end{eqnarray*}
since $\sum_{h=1}^{H_{b}} S_{(i,b)}^h \geq L_{b}.$ The rest of the proof is same as the proof of
Lemma \ref{lem: norelay}.
\end{proof}

\begin{thm} Consider the two-dimensional i.i.d. mobility model with slow mobiles, and assume that Assumption 1-3 holds.
Then given a delay constraint $D,$ we have
\begin{eqnarray}
\frac{4\sqrt[3]{2}WT}{\sqrt[3/2]{\Delta}}\sqrt[3/2]{n}\left(\sqrt[3]{D}+1\right)\geq
E[{\Lambda}[T]]. \label{eq: nrt_s}
\end{eqnarray}
\label{thm: upper_s}
\end{thm}
\begin{proof}Similar to the proof of Theorem \ref{thm: upper_d}.
\end{proof}

From Theorem \ref{thm: upper_s}, we can conclude that the throughput per S-D is $O(\sqrt[3]{D/n})$
given a delay constraint $D.$

\subsection{Joint Coding-Scheduling Algorithm}

In this subsection, we propose a joint coding-scheduling scheme to achieve the throughput suggested
in Theorem \ref{thm: upper_s}. In the receiving step, we divide the unit square into square cells
with each side of length equal to $1/\sqrt[6]{nD^2},$ which is of the same order as the optimal
hitting distance obtained in Section \ref{sec: K}. The mean number of nodes in each cell will be
denoted by $M_2$ and is equal to $\sqrt[3/2]{n/D}.$ The packet size is chosen to be
$$\frac{10W}{11c_s C \sqrt{M_2}}$$ so that at each time slot, all nodes in a good cell
can transmit one packet to some other node in the same cell by using the highway algorithm proposed
in \cite{FraDouTseThi_05} (see in Appendix B), where $c_s$ is a constant independent of $n.$  In
the broadcasting step, the unit square is divided into square cells with each side of length equal
to $1/\sqrt[6]{n^2 D}.$ The mean number of nodes in each will be denoted by $M_1$ and is equal to
$\sqrt[3]{n/D}.$ In the broadcasting step, the transmission radius of each nodes is chosen to be
$\sqrt{2}\sqrt[6]{n^2 D}.$ Note the packet size is
$$\frac{10W}{11c_s C \sqrt{M_2}}=\frac{10W}{11c_s C M_1}.$$
So in the broadcasting step, all nodes in a good cell could be scheduled to broadcast one coded
packet at one min-slot. Also note that $M_1M_2D/n = 1.$

\noindent{\bf Joint Coding-Scheduling Algorithm II:} We group every $16D$ time slots into a super
time slot. At each super time slot, the nodes transmit packets as follows:

\begin{enumerate}
\item[(1)] {\bf Raptor Encoding:} Each source takes $2D/5$ data packets, and uses Raptor codes to
generate $D$ coded packets.

\item[(2)]  {\bf Broadcasting:} The unit square is divided into a regular lattice with $n/M_1$
cells. This step consists of $D$ time slots. At each time slot, the nodes execute the following
tasks:
\begin{enumerate}
\item[(i)] In each good cell, the nodes take their turns to broadcast a coded packet to $9M_1/10$
other nodes in the cell. We use the same definition of  a good cell as in Algorithm I, i.e., the
number of nodes in a good cell should be with a factor of the mean, where the factor is required to
lie in the interval $[0.9, 1.1].$

\item[(ii)] All nodes check the duplicate packets they have. If more than one duplicate packet is
destined to a same destination, randomly keep one and drop the others.
\end{enumerate}

\item[(3)]{\bf Receiving:} The unit square is divided into a regular lattice with $n/M_2$ cells.
This step consists of $15D$ time slots. At each time slot, the nodes in a good cell execute the
following tasks in the mini-slot allocated to that cell.

\begin{enumerate}
\item[(i)] Each node containing deliverable packets randomly selects a deliverable packet, and
sends a request to the corresponding destination.

\item[(ii)] Each destination only accepts one request and refuses the others.

\item[(iii)] The nodes whose requests are accepted transmit the deliverable packets to their
destinations using the highway algorithm proposed in \cite{FraDouTseThi_05}.
\end{enumerate} At the end of this step, all undelivered packets are dropped. Destinations use Raptor decoding to obtain the source
packets. Note that one requires some overhead in obtaining route to the destination to perform step
(3)(i) above. As in previous works, we assume that this overhead is small since one can transmit
many packets in each time slot, under the slow mobility assumption.
\end{enumerate}

\begin{thm}
Consider Joint Coding-Scheduling Algorithm II. Suppose $D$ is both $\omega(1)$ and $o(n),$ and the
delay constraint is $16D.$ Then given any $\epsilon,$ there exists $n_0$ such that for any $n\geq
n_0,$ every data packet sent out can be recovered at the destination with probability at least
$1-\epsilon,$ and furthermore
\begin{eqnarray}\lim_{T\rightarrow \infty} \Pr\left(\frac{\Lambda_i[T]}{T}\geq
\left(\frac{9W}{440 c_s C}\right)\sqrt[3]{\frac{D}{n}} \hbox{ } \forall i\right)=1.\label{eq: th_s}
\end{eqnarray}
\label{thm: TSSM}
\end{thm}
\begin{proof}
The proof will show that the following events happen with high probability.

\noindent{\bf Broadcasting:} At least $4D/5$ coded packets from a source are successfully
duplicated after the broadcasting step with high probability, i.e.,
\begin{eqnarray}\Pr\left(A_i[t_s]\geq \frac{4}{5}D\right)\geq
1-3e^{-\frac{D}{100000}}, \label{eq: cdt1}\end{eqnarray} where a coded packet is said to be
successfully duplicated if the packet is in $4M_1/5$ distinct relay nodes.

\noindent{\bf Receiving:} At least $D/2$ distinct coded packets from a source are delivered to its
destination after the receiving step with high probability, i.e.,
\begin{eqnarray}
\Pr\left(\left.B_i[t_s]\geq \frac{D}{2}\right| A_i[t_s]\geq \frac{4}{5}D\right)\geq
1-e^{-\frac{D}{5000}}.\label{eq: cdt2}
\end{eqnarray}

After obtaining inequality (\ref{eq: cdt1}) and (\ref{eq: cdt2}), the theorem can be proved by
following the argument in Theorem \ref{thm: ESTM}.

{\bf Analysis of broadcasting:} Similar to the analysis of inequality (\ref{eq: cdt1_2}).

{\bf Analysis of receiving:} Assume that coded packets $(i,1), \ldots, \left(i, 4D/5\right)$ are
successfully duplicated. Note that duplicate packets from a common source are carried by different
nodes; and the mobilities of those nodes are independent. So $(i,k)$ will be definitely delivered
to its destination if both the following conditions hold:
\begin{enumerate}
\item[(i)] A copy of $(i,k)$ is the only deliverable packet for destination $i+1$ in time slot $t.$
Let ${\cal D}^1_{(i,k)}[t]$ denote this event; and assume that the duplicate packet is in node $j.$

\item[(ii)] Node $j$ has no other deliverable packet, and the cell containing node $j$ is good. Let
${\cal D}^2_{(i,k)}[t]$ denote this event.
\end{enumerate}

Let $\tilde{M}_{(i,k)}$ denote the number of copies of $(i,k).$ Since each source has at most
$9M_1D/10$ duplicate packets in the network, we have
\begin{eqnarray*}
\Pr\left({{\cal D}^1_{(i,k)}[t]}\right)\geq \frac{\tilde{M}_{(i,k)} M_2}{n}
\left(1-\frac{M_2}{n}\right)^{\frac{9}{10}M_1D-1}.
\end{eqnarray*}
It is easy to verify that $$\lim_{n\rightarrow
\infty}\left(1-\frac{M_2}{n}\right)^{\frac{9}{10}M_1D-1}=e^{-0.9}.$$ Thus, for successfully
duplicated packet $(i,k),$ i.e., $\tilde{M}_{(i,k)}\geq 4M_1/5,$ we can conclude that
\begin{eqnarray}
\Pr\left({{\cal D}^1_{(i,k)}[t]}\right)\geq \frac{4M_1 M_2}{5 e n } \label{eq: p1}
\end{eqnarray} holds for sufficiently large $n$. Note that each node carries at most $DM_1$ duplicate packets,
so we further have
\begin{eqnarray}
\Pr\left(\left.{{\cal D}^2_{(i,k)}[t]}\right|{{\cal D}^1_{(i,k)}[t]}\right)\geq
\left(1-\frac{M_2}{n}\right)^{DM_1-1}-2e^{-\frac{M_2}{300}}, \label{eq: p2}
\end{eqnarray} where $(1-M_2/n)^{DM_1-1}$ is the lower bound on the probability that all packets
in node $j$ except $(i,k,j)$ are undeliverable, and $1-2e^{-\frac{M_2}{300}}$ is the probability
that the cell is good.

From inequalities (\ref{eq: p1}) and (\ref{eq: p2}), we can conclude that for sufficiently large
$n,$
\begin{eqnarray*} &&\Pr\left({{\cal D}^1_{(i,k)}[t]},
{{\cal D}^2_{(i,k)}[t]}\right)\\
&=&\Pr\left({{\cal D}^1_{(i,k)}[t]}\right)\Pr\left(\left.{{\cal D}^2_{(i,k)}[t]}\right|
{{\cal D}^1_{(i,k)}}[t]\right)\\
&\geq&\frac{M_1 M_2}{12 n }\\
&=&\frac{1}{12D}.
\end{eqnarray*}
Inequality (\ref{eq: cdt2}) can be proved by the balls-and-bins argument used to show inequality
(\ref{eq: cdt2_2}).
\end{proof}

\section{Conclusion}
\label{sec: conl} In this paper, we investigated the optimal delay-throughput trade-off in ad-hoc
networks with two-dimensional i.i.d. mobility models. For the two-dimensional i.i.d. mobility model
with fast mobiles, the optimal trade-off was shown to be $\lambda=\Theta\left(\sqrt{D/n}\right)$
when $D$ is both $\omega(\sqrt[3]{n})$ and $o(n).$ For the two-dimensional i.i.d. mobility model
with slow mobiles, the optimal trade-off was shown to be $\lambda=\Theta\left(\sqrt[3]{D/n}\right)$
when $D$ is both $\omega(1)$ and $o(n).$

We now briefly comment on the conditions that we have imposed on the delay requirement to obtain
the optimal delay-throughput tradeoffs. In the slow mobility case, we have assumed that the
required delay has to be $\omega(1).$ This condition on the delay is used to allow the decoding
error probability to go to zero as $n\rightarrow \infty.$ If we allow a small probability of loss
(it can be arbitrarily small), then one can allow the delay to be $\Theta(1).$ We have also assumed
that the delay is $o(n).$ This is not really a restriction since it is easy to see from prior work
that the best achievable throughput of $\Theta(1)$ is obtained when the delay is $\omega(n)$
\cite{GroTse_01}. The $o(n)$ condition on delay is used in our paper only to ensure that our cell
partitioning, scheduling and coding strategy works.

In the case of fast mobiles, when $D$ is $O(\sqrt[3]{n}),$ then the number of packets that can be
transmitted in $D$ time slots is a constant and hence one cannot use coding to ensure that the
probability of packet loss is arbitrarily small. In this case, one can obtain a bound that is a
logarithmic factor smaller than the upper bound using packet replication techniques as has been
done in \cite{LinShr_04} for the slow-mobile case. However, the best achievable lower bound is
unknown. Again the $o(n)$ requirement is not significant since a throughput of $\Theta(1)$ can be
achieved if the delay requirement is larger \cite{GroTse_01}.

\textbf{Acknowledgment:} The authors gratefully acknowledge the useful discussions with Prof. Geir
E. Dullerud and Prof. Bruce Hajek, University of Illinois at Urbana-Champaign. We also thank Prof.
Michael J. Neely, University of Southern California, for his comments on the earlier version of
this paper.

\section*{Appendix A: Raptor Codes}

\subsection{Raptor Codes} Raptor codes are low-complexity, near-optimal rate-less codes for erasure channels. It was proposed by Shokrollahi in \cite{Sho_04},
and the following result was presented in \cite{Sho_04}.

\begin{lemma}
The receiver can correctly decode the $M$ data packets with probability at least
$1-1/(M^{a(\epsilon)})$ for some $a(\epsilon)>0$ after it obtains $(1+\epsilon)M$ coded packets
generated by Raptor codes. The number of operations used for encoding and decoding is $O(M).$
\label{lem: RC}

\rightline{$\square$}
\end{lemma}

\section*{Appendix B: Throughput of Static Wireless Networks} The throughput of a random wireless network with
$n$ static nodes and $n$ random S-D pairs is introduced by Gupta and Kumar \cite{gupkum00}. They
showed that the maximum throughput per S-D pair is $O(1/\sqrt{n}),$ and proposed a scheduling
scheme achieving a throughput of $\Theta(1/\sqrt{n\log n})$ per S-D pair. This $\log n$ gap was
latter closed by Franceschetti et. al in \cite{FraDouTseThi_05} where they showed a throughput of
$\Theta(1/\sqrt{n})$ per S-D pair is achievable. The result is obtained under the physical
interference models. However, it can be easily extended to the protocol model by using the same
algorithm.
\begin{lemma}
In a random wireless network with $n$ static nodes and $n$ S-D pairs, a throughput of
$$\lambda=\frac{W}{c_s\sqrt{n}}$$ bits/time-slot per S-D pair is achievable, where $c_s$ is a positive constant independent of $n.$  \label{thm: Tse}
\label{lem: Tse} \rightline{$\square$}
\end{lemma}

Suppose the nodes use a common transmission radius $r=\Theta(1/n).$ The key idea of
\cite{FraDouTseThi_05} is to construct $\Theta(n)$ disjoint paths traversing the network vertically
and horizontally. These paths are called highways in \cite{FraDouTseThi_05}, and a throughput of
$\Theta(1/\sqrt{n})$ per S-D pair is achievable by transmitting data throughput these highways. We
call this algorithm a highway algorithm in this paper.

\section*{Appendix C: Probability Results}
In this appendix, we present some standard results in probability for the reader's convenience. In
addition, we also present some variations of standard results which do not seem to be available in
any book to best of our knowledge.

The following lemma is a standard result in probability, which we provide here for convenience.
\begin{lemma}
Let $X_1,\ldots, X_n$ be independent $0-1$ random variables such that $\sum_i X_i=\mu.$ Then, the
following Chernoff bounds hold
\begin{eqnarray}
\hbox{Pr}\left(\sum_{i=1}^n X_i<(1-\delta)\mu\right)&\leq & e^{-\delta^2\mu/2};\label{eq: ChB_1}\\
\hbox{Pr}\left(\sum_{i=1}^n X_i>(1+\delta)\mu\right)& \leq &e^{-\delta^2\mu/3}.\label{eq: ChB_2}
\end{eqnarray}
\label{lem: C_B}
\end{lemma}
\begin{proof}
A detailed proof can be found in \cite{MitUpf_05}.
\end{proof}

The next lemmas are variations of standard balls-and-bins problems. However, we have not seen the
results for the particular variation that we need in this paper. So we present the lemmas along
with brief proofs below.
\begin{lemma}
Assume we have $m$ bins. At each time, choose $h$ bins and drop one ball in each of them. Repeat
this $n$ times. Using $N_1$ to denote the number of bins containing at least one ball, the
following inequality holds for sufficiently large $n.$
\begin{eqnarray}
\hbox{Pr}\left(N_1\leq (1-\delta) m\tilde{p}_1 \right)&\leq& 2e^{-\delta^2
m\tilde{p}_1/3}.\label{eq: BB_2}
\end{eqnarray}
where $\tilde{p}_1=1-e^{-\frac{nh}{m}}.$ \label{lem: ball-bin_2}
\end{lemma}
\begin{proof}
At each time, bin $i$ receives a ball with probability $h/m.$ We let $\kappa_i$ denote the number
of balls in bin $i.$ Now consider a related balls-and-bins problem where the ball dropping
procedure is replaced by a certain number of trials as dictated by a Poisson random variable.
Specifically, define $\tilde{n}$ to be a Poisson random variable with mean $n,$ and repeat the ball
dropping procedure $\tilde{n}$ times. Let $\tilde{\kappa}_i$ denote the number of balls in bin $i$
in this case. It is easy to see that $\{\tilde{\kappa}_i\}$ are i.i.d. Poisson random variables
with mean $nh/m.$ So we can conclude
\begin{eqnarray*}
\Pr\left(N_1\leq (1-\delta)m \tilde{p}_1\right)
&=&\Pr\left(\sum_{i=1}^m 1_{\kappa_i\geq 1}\leq  (1-\delta)m \tilde{p}_1 \right)\\
&\leq & \Pr\left(\left.\sum_{i=1}^m 1_{\tilde{\kappa}_i\geq 1} \leq (1-\delta)m \tilde{p}_1 \right| \tilde{n}\geq n \right)\\
&\leq&\frac{\hbox{Pr}\left(\sum_{i=1}^m 1_{\tilde{\kappa}_i\geq 1} \leq (1-\delta) m \tilde{p}_1
\right)}{\hbox{Pr}(\tilde{n}\geq n)}\\
&=&2 \hbox{Pr}\left(\sum_{i=1}^m 1_{\tilde{\kappa}_i\geq 1} \leq (1-\delta) m \tilde{p}_1 \right).
\end{eqnarray*}
Since
\begin{eqnarray*}
\hbox{Pr}\left(1_{\tilde{\kappa}_i\geq 1}=1\right)=\hbox{Pr}\left(\tilde{\kappa}_i\geq
1\right)=1-e^{-\frac{nh}{m}}=\tilde{p}_1,
\end{eqnarray*}
from Lemma \ref{lem: C_B}, we have
\begin{eqnarray*}
\Pr\left(N_1\leq (1-\delta) m \tilde{p}_1 \right) &\leq& 2 e^{-\delta^2 m\tilde{p}_1/3}.
\end{eqnarray*}
\end{proof}

The above idea of using a Poisson number of trials to bound the probability of the occurrence of an
event in a fixed number of trials is called the Poisson heuristic in \cite{MitUpf_05}.

\begin{lemma}
Suppose $n$ balls are independently dropped into $m$ bins and one trash can. After a ball is
dropped, the probability in the trash can is $1-p,$ and the probability in a specific bin is $p/m.$
Using $N_2$ to denote the number of bins containing at least $1$ ball, the following inequality
holds for sufficiently large $n.$
\begin{eqnarray}
\hbox{Pr}\left(N_2\leq (1-\delta) m \tilde{p}_2\right)&\leq& 2e^{-\delta^2 m
\tilde{p}_2/3};\label{eq: BB_1}
\end{eqnarray}
where $\tilde{p}_2=1-e^{-\frac{np}{m}}.$ \label{lem: ball-bin}
\end{lemma}
\begin{proof}
Let $\kappa_i$ denote the number of balls in bin $i.$ Next define $\tilde{n}$ to be a poisson
random variable with mean $n.$ We consider the case such that $\tilde{n}$ balls are independently
dropped in $m$ bins. Using $\tilde{\kappa}_i$ to be number of balls in bin $i$ in this case, it is
easy to see that $\{\tilde{\kappa}_i\}$ are i.i.d. poisson random variables with mean
$\frac{np}{m}.$

Now given $n_b,$ we first have
\begin{eqnarray*}
\Pr\left(N_2\leq (1-\delta)m \tilde{p}_2\right)&=&\Pr\left(\sum_{i=1}^m 1_{{\kappa}_i\geq 1}\leq (1-\delta)m \tilde{p}_2\right)\\
&=& \Pr\left(\left.\sum_{i=1}^m 1_{\tilde{\kappa}_i\geq 1}\leq (1-\delta) m \tilde{p}_2 \right| \tilde{n}\geq n \right)\\
&\leq&\frac{\hbox{Pr}\left(\sum_{i=1}^m 1_{\tilde{\kappa}_i\geq 1} \leq (1-\delta) m \tilde{p}_2
\right)}{\hbox{Pr}(\tilde{n}\geq n)}.
\end{eqnarray*}
Since
\begin{eqnarray*}
\hbox{Pr}\left(1_{\tilde{\kappa}_i\geq 1}=1\right)=\hbox{Pr}\left(\tilde{\kappa}_i\geq
1\right)=1-e^{-\frac{n p}{m}}=\tilde{p}_2,
\end{eqnarray*} from Lemma \ref{lem: C_B}, we have
\begin{eqnarray*}
\hbox{Pr}\left(\sum_{i=1}^m 1_{\tilde{\kappa}_i\geq 1} \leq (1-\delta) m \tilde{p}_2 \right)\leq
e^{-\delta^2 m \tilde{p}_2/3 }.
\end{eqnarray*}
which implies for sufficiently large $n,$
\begin{eqnarray*}
\Pr\left(N_2\leq (1-\delta)m \tilde{p}_2\right)&\leq& \sqrt{3\pi n }e^{-\delta^2 m \tilde{p}_2/3}\\
&\leq&2 e^{-\delta^2 m \tilde{p}_2/3}.
\end{eqnarray*}
\end{proof}
\end{document}